\documentclass[a4paper,11pt]{article}

\usepackage{jheppub} 
\usepackage[T1]{fontenc}
\usepackage{graphicx}

\newcommand{\bg}{\mbox{\boldmath $\gamma$}}
\newcommand{\bt}{\mbox{\boldmath $\theta$}}

\newcommand{\by}{\mbox{\boldmath $y$}}

\title{\boldmath Hessian geometry and entanglement thermodynamics}

\author{Hiroaki Matsueda}
 
\affiliation{Sendai National College of Technology, Sendai 989-3128, Japan}

\emailAdd{matsueda@sendai-nct.ac.jp}

\abstract{
We reconstruct entanglement thermodynamics by means of Hessian geometry, since this method exactly generalizes thermodynamics into much wider exponential family cases including quantum entanglement. Starting with the correct first law of entanglement thermodynamics, we derive that a proper choice of the Hessian potential leads to both of the entanglement entropy scaling for quantum critical systems and hyperbolic metric (or AdS space with imaginary time). We also derive geometric representation of the entanglement entropy in which the entropy is described as integration of local conserved current of information flowing across an entangling surface. We find that the entangling surface is equivalent to the domain boundary of the Hessian potential. This feature originates in a special property of critical systems in which we can identify the entanglement entropy with the Hessian potential after the second derivative by the canonical parameters, and this identification guarantees violation of extensive nature of the entropy.
}

\keywords{Hessian geometry, entanglement thermodynamics, AdS/CFT correspondence}

\begin{document}

\maketitle

\flushbottom

\section{Introduction}

In my recent paper~\cite{Matsueda1}, we have developed a theory for information-geometrical interpretation of the anti-de Sitter space/conformal field theory (AdS/CFT) correspondence~\cite{Maldacena1,Maldacena2}. In particular, we have focused on how the information of the reduced density matrix for free fermions is rigorously mapped onto the canonical-parameter space. Since the reduced density matrix has thermal properties owing to partial truncation of environmental degrees of freedom, the density-matrix eigenvalue has an exponential family form like a thermal distribution. Then, the Fisher information metric has Hessian structure that provides us with beautiful geometric properties. We have found that the precise determination of a functional form of the Hessian potential and the nontrivial mapping from original model parameters onto the canonical parameters are two crucial things to find rigorous correspondence between AdS and CFT. As we have commented in the previous paper, it seems quite interesting to examine more about Hessian geometry by means of exact reconstruction of entanglement thermodynamics~\cite{Blanco,Casini,Wong,Takayanagi2,Takayanagi3,Faulkner,Banerjee,Nima}. The aim of this paper is to addresses this issue.

We emphasize that the entanglement thermodynamics by the present approach contains more information rather than simple entropy-energy relation frequently used in many literatures~\cite{Blanco,Casini,Wong,Takayanagi2,Takayanagi3,Faulkner,Banerjee,Nima}. This is because our method naturally extends standard thermodynamics into general exponential family cases including quantum entanglement, and emables us to compare the entanglement thermodynamics with the standard one in detail. Simply speaking, the entropy-energy relation is constructed by the first variation of the entanglement entropy. However, we can access background information before taking the variation by considering the exponential family form, and thus this method is quite powerful. Furthermore, in the present approach, the entanglement entropy is directly connected to the Fisher metric, and it is thus straightforward to find geometric representation of the entanglement entropy. Since the examination of the entropy-energy relation is aimed for deeper understanding of holographic entanglement entropy, it is quite meaningful to develop a holographic description of the entanglement entropy in term of the present approach as well as general construction of the entanglement thermodynamics. More precisely speaking, the central piece is the Hessian potential, and the Fisher metric is exactly defined by the second derivative of the Hessian potential by the canonical parameters. In some cases, the entanglement entropy may be identified with the Hessian potential after the second derivative by the canonical parameters. This special situation actually occurs, when we consider the entanglement entropy of CFT. This situation is completely different from the standard thermodynamics, and actually provides us with violation of extensive nature of the entropy.

In the geometric representation of the entanglement entropy, the canonical parameters for the exponential family form can be ragarded as local conserved current of information that flows across an entangling surface, and their integration is equivalent to the entanglement entropy. We will find that the entangling surface is equivalent to the domain boundary of the Hessian potential. Therefore, the mathematical structure of the Hessian potential is a key for full construction of the physics of entanglement thermodynamics in the holographic side. We should note again that the identification between the potential and the entropy after the second derivative is a very special nature that the quantum critical models have, and this idenfication guarantees violation of extensive nature of the entropy.

The organization of this paper is as follows. In Sec.~II, we construct a general theory for entanglement thermodynamics by based on the Hessian geometry. In Sec.~III, we discuss about geometric representation of the entanglement entropy as information flow across an entangling surface, and will prove that the surface is equivalent with the domain boundary of the Hessian potential. The final section is devoted to the summary part.

\section{General construction of entanglement thermodynamics based on Hessian geometry}

\subsection{Equivalence between entanglement spectrum and Hessian structure}

Let us start with the ground state $\left|\psi\right>$ for a given Hamiltonian $H_{A+\bar{A}}$ in $d$-dimensional (dD) flat Minkowski spacetime $R^{1,d}$ ($d$ is a space dimension),
\begin{eqnarray}
H_{A+\bar{A}}\left|\psi\right>=E_{0}\left|\psi\right>,
\end{eqnarray}
where $E_{0}$ is the ground-state energy. Here, the total system is devided into a subsystem $A$ with size $L$ and it's complement $\bar{A}$. The size $L$ has been regularized by a lattice constant. Then, the ground state is described by the Schmidt decomposition,
\begin{eqnarray}
\left|\psi\right>=\sum_{n}\sqrt{\lambda_{n}}\left|A;n\right>\otimes\left|\bar{A};n\right>,
\end{eqnarray}
where $\left|A;n\right>$ and $\left|\bar{A};n\right>$ are Schmidt bases of $A$ and $\bar{A}$, respectively, and $\sqrt{\lambda_{n}}$ is the Schmidt coefficient with a semipositive-definite $\lambda_{n}$. We focus on the partial density matrix, $\rho_{A}={\rm tr}_{\bar{A}}\left|\psi\right>\left<\psi\right|$. More explicitely, $\rho_{A}$ is described as
\begin{eqnarray}
\rho_{A}=\sum_{n}\lambda_{n}\left|A;n\right>\left<A;n\right|.
\end{eqnarray}
where the Schmidt coefficient is normalized as $\sum_{n}\lambda_{n}=1$. Thus, the partial density matrix has structure similar to a thermal distribution, even though we consider the ground state. If we define $\lambda_{n}=Z^{-1}e^{-\epsilon_{n}}=\exp\left(-\epsilon_{n}-\ln Z\right)$, The operator form of $\rho_{A}$ is given by
\begin{eqnarray}
\rho_{A} = \frac{1}{Z}e^{-\tilde{H}_{A}},
\end{eqnarray}
where $\epsilon_{n}$ is the eigenvalue of $\tilde{H}_{A}$ and $Z=\sum_{n}e^{-\epsilon_{n}}$. The eigenvalue is called the entanglement energy. The operator $\tilde{H}_{A}$ is called the entanglement Hamiltonian and is in general different from the Hamiltonian in the original quantum system, $H_{A+\bar{A}}$.

To examine the feature of the density-matrix eigenvalue $\lambda_{n}$ in more detail, we consider a 1D free fermion model. We notice that the eigenvalue $\lambda_{n}$ is a function of partial system size of $A$, $L$, filling fraction of fermions, $\delta$, and time after some perturbation to the ground state, $t$. Then, there exist a parameter set $\bt=(\theta^{1},\theta^{2},\theta^{3})$ that depends on $L$, $\delta$, and $t$, i.e. $\theta^{\alpha}=\theta^{\alpha}(L,\delta,t)$ for $\alpha=1,2,3$, and we suppose that the eigenvalue $\lambda_{n}$ has the following exponential family form,
\begin{eqnarray}
\lambda_{n}(\bt) = \exp\left\{\theta^{\alpha}F_{n\alpha}-\psi(\bt)\right\}, \label{expo}
\end{eqnarray}
where $\bt$ is called the canonical parameter. In other words, the above statement requires that the entanglement energy has a covariant form $\epsilon_{n}=-\theta^{\alpha}F_{n\alpha}$ and the potential function is defined by $\psi(\bt)=\ln Z$. Actually, in my previous paper~\cite{Matsueda1}, we have found the following mapping,
\begin{eqnarray}
\bt=\left(\theta^{1},\theta^{2},\theta^{3}\right)=\left(\frac{1}{L^{2}},\frac{h(\delta)}{L},\frac{t}{L}\right), \label{t123}
\end{eqnarray}
where $h$ is a function of filling $\delta$ with particle-hole symmetry. Hereafter, we examine geometric representation of Eq.~(\ref{expo}) in a sense that we construct a geometric space spanned by the canonical parameter $\bt$ with the help of the Fisher information metric.

In general, we can consider $D$-dimensional space spanned by the $D$-canonical parameters. It is an interesting but still open question whether the condition $D=(d+1)+1$ holds for arbitrary $d$ and the additional degree of freedom originates in a kind of a renormalization scale. In a 1D case, this is satisfied as we have examined in the previous paper.

For later convenience, we abbreviate the expectation value of a function $O_{n}(\bt)$ by the angle bracket as $\left<\mbox{\boldmath $O$}\right>=\sum_{n}\lambda_{n}(\bt)O_{n}(\bt)$, where we omit the index $\bt$ in the bracket and use the bold symbol. By defining the entanglement spectrum as
\begin{eqnarray}
\gamma_{n}(\bt)=-\ln\lambda_{n}(\bt), \label{gamma}
\end{eqnarray}
the entanglement entroy is given by
\begin{eqnarray}
S(\bt) = -\sum_{n}\lambda_{n}(\bt)\ln\lambda_{n}(\bt) = \left<\bg\right>. \label{entropy}
\end{eqnarray}
The entanglement entropy is one of key parameters throughout this paper. The geometry we consider is constructed by the Fisher metric defined by~\cite{Amari}
\begin{eqnarray}
g_{\mu\nu}(\bt)=\sum_{n}\lambda_{n}(\bt)\frac{\partial\gamma_{n}(\bt)}{\partial\theta^{\mu}}\frac{\partial\gamma_{n}(\bt)}{\partial\theta^{\nu}}=\left<\partial_{\mu}\bg\partial_{\nu}\bg\right>. \label{Fisher}
\end{eqnarray}
This is a sort of relative entanglement entropy. Thus, we are going to measure physical difference of two similar quantum states. Note that the diagonal parts of the metric are all positive. Therefore, the positive sign property indicates that the Lorentzian signature does not appear and we need to take imaginary time.

The Fisher metric has an another form. An important equality is
\begin{eqnarray}
\left<\partial_{\nu}\bg\right>=0. \label{gn}
\end{eqnarray}
Differentiating this equality by $\theta^{\mu}$ leads to
\begin{eqnarray}
g_{\mu\nu}=\left<(\partial_{\mu}\bg)(\partial_{\nu}\bg)\right>=\left<\partial_{\mu}\partial_{\nu}\bg\right>. \label{Fisher2}
\end{eqnarray}
Thus we have two different representations of the Fisher metric. Fortunately, the corresponding geometry becomes quite simple in the second representation of Eq.~(\ref{Fisher2}), and this is the so-called Hessian geometry~\cite{Shima}. Since the entanglement spectrum $\gamma_{n}(\bt)$ has been given by
\begin{eqnarray}
\bg_{n}(\bt)=\psi(\bt)-\theta^{\alpha}F_{n\alpha}, \label{gamma2}
\end{eqnarray}
the second derivative of this equation by $\bt$ leads to
\begin{eqnarray}
g_{\mu\nu} = \left<\partial_{\mu}\partial_{\nu}\bg\right>=\partial_{\mu}\partial_{\nu}\psi(\bt). \label{rep1}
\end{eqnarray}
This is the so-called Hessian structure. The Fisher metric is characterized by $\psi(\bt)$, and $\psi(\bt)$ is called the Hessian potential.

\subsection{First law of entanglement thermodynamics}

Next we consider the entanglement entropy to examine information-geometrical notions of fundamental laws of entanglement thermodynamics such as the entropy-energy relation. Taking the statistical avarage of Eq.~(\ref{gamma2}), we find
\begin{eqnarray}
S(\bt)=\psi(\bt)-\theta^{\alpha}\left<F_{\alpha}\right>. \label{sav}
\end{eqnarray}
To evaluate $\left<F_{\alpha}\right>$, we calculate the first derivative of $\bg$ by $\bt$ as
\begin{eqnarray}
\partial_{\mu}\bg = \partial_{\mu}\psi(\bt)-F_{n\mu}, \label{partial1}
\end{eqnarray}
and taking the average of Eq.~(\ref{partial1}) with the help of Eq.~(\ref{gn}), we find 
\begin{eqnarray}
\left<F_{\mu}\right>=\partial_{\mu}\psi(\bt)=\eta_{\mu}(\bt), \label{fav}
\end{eqnarray}
where $\eta_{\alpha}$ is called as the Legendre parameter. Combining Eq.~(\ref{sav}) with Eq.~(\ref{fav}), we can rewrite the entropy as
\begin{eqnarray}
S(\bt) = \psi(\bt)-\theta^{\alpha}\partial_{\alpha}\psi(\bt). \label{solve}
\end{eqnarray}
The first derivative of the entropy is represented as
\begin{eqnarray}
\partial_{\nu}S(\bt)=-\theta^{\alpha}\partial_{\alpha}\partial_{\nu}\psi(\bt)=-\theta^{\alpha}g_{\alpha\nu}. \label{dS}
\end{eqnarray}
These two relations are fundamentals of entanglement thermodynamics. When we introduce the entanglement temperature $T_{E}$, we find the generalized first law of thermodynamics,
\begin{eqnarray}
F = E - T_{E}S,
\end{eqnarray}
where we have defined
\begin{eqnarray}
F&=&-T_{E}\psi=-T_{E}\ln Z, \\
E&=&-T_{E}\theta^{\alpha}\eta_{\alpha}.
\end{eqnarray}

These results are extremely important in terms of AdS/CFT, since the entropy is directly related to the spacetime metric as shown in Eq.~(\ref{dS}). This enable us to perform quite simple geometric description of entanglement thermodynamics. The first law of entanglement thermodynamics suggests that the Hessian potential can be identified with the entanglement entropy after the second derivative. Let us consider a situation where the entropy is a logarithmic function of $\theta^{1}$, $S(\bt)=A\ln\theta^{1}$. This situation corresponds to 1D quantum critical systems, since $\theta^{1}=L^{-2}$ and then the logarithmic entanglement-entropy scaling appears. In this case, the general solution of Eq.~(\ref{solve}) is
\begin{eqnarray}
\psi(\bt) = A\ln\theta^{1}+A+\theta^{\alpha}F_{0\alpha} = S(\bt)+A+\theta^{\alpha}F_{0\alpha}.
\end{eqnarray}
This leads to identification between $S$ and $\psi$ after the second derivative by the canonical parameters. Thus, we propose
\begin{eqnarray}
g_{\mu\nu}(\bt)=\partial_{\mu}\partial_{\nu}\psi(\bt)=\partial_{\mu}\partial_{\nu}S(\bt).
\end{eqnarray}
When $\theta^{1}$ corresponds to a kind of length scale, the feature of this identification holds even in higher-dimensional cases where the entanglement-entropy scaling is given by the area law, not a logarithmic function. We will later mention this point. Because of these properties, we focus on the properties of Hessian potential in CFT instead of entanglement entropy itself.

\subsection{Derivation of hyperbolic metric from Hessian potential for CFT}

Let us examine the explicit functional form of the Hessian potential that provides us with some essence of the AdS/CFT correspondence. Suppose the following potential function
\begin{eqnarray}
\psi(\bt)=-\kappa\ln f=-\kappa\ln\left\{ \theta^{1}-\frac{1}{2}\sum_{i=2}^{D}\left(\theta^{i}\right)^{2} \right\}, \label{fullpotential}
\end{eqnarray}
with a positive constant $\kappa$. The domain of this potential function is
\begin{eqnarray}
\theta^{1}>(1/2)\sum_{i=2}^{D}(\theta^{i})^{2}. \label{domain}
\end{eqnarray}
This functional form is applicable to arbitrary $D$ cases. The assumptions for the functional form and the domain are reasonable for our choice of the parameters in Eq.~(\ref{t123}) for 1D free fermions. Substituting Eq.~(\ref{t123}) into Eq.~(\ref{domain}), we obtain
\begin{eqnarray}
\left(\frac{1}{L}\right)^{2} > \frac{1}{2}\left\{ \left(\frac{h(\delta)}{L}\right)^{2}+\left(\frac{t}{L}\right)^{2} \right\}.
\end{eqnarray}
If we assume $t<L$, this condition is satisfied. We then find that the potential in Eq.~(\ref{fullpotential}) is expanded as
\begin{eqnarray}
\psi(L,\delta,t) \simeq 2\kappa\ln L + \frac{1}{2}\kappa\left\{ h^{2}(\delta) + t^{2} \right\}.
\end{eqnarray}
Since we can identify $\psi(L,\delta,t)$ with $S(L,\delta,t)$, this is actually consistent with the entanglement entropy scaling of 1D critical systems for $\kappa=c/6$ with the central charge $c$~\cite{Holzhey,Calabrese1,Calabrese2,Calabrese3,AAAL,Liu,Nezhadhaghighi,Hubeny,Brown}.

In spatially 2D cases ($d=2$) with linear system size $L$, we guess from our numerical experiences that $\theta^{1}$ is a kind of scale parameter and $\theta^{1}$  behaves as
\begin{eqnarray}
\theta^{1}\simeq e^{-aL/\kappa}, \label{2D}
\end{eqnarray}
and then the area law formula can be derived, $\psi(\bt)\sim aL=aL^{d-1}$. This means that the density of pseudo-energy levels in the entanglement Hamiltonian is extremely higher than that in 1D cases. We think that this is reasonable situation, and that is one reason why the density matrix renormalization group calculation is hard in 2D cases inspite of extreme powerfulness in 1D cases. In this 2D case also, the entropy is given by a logarithmic function for a scale parameter $\theta^{1}$. Thus, the feature of the identification between $\psi$ and $S$ is kept. This point is what we have discussed in the last part in the previous subsection.

Let us also look at the classical side. For the full potential function $\psi(\bt)$ in Eq.~(\ref{fullpotential}), there exists a parameter set $\by=\left(y^{1},y^{2},...,y^{D}\right)$ for which the metric tensor is exactly the hyperbolic form in the Poincare disk representation. For this proof, it is helpful to introduce the Legendre transformation $\eta_{\alpha}=\partial_{\alpha}\psi$. Then, the metric is represented as
\begin{eqnarray}
g_{\mu\nu}=\partial_{\mu}\partial_{\nu}\psi=\partial_{\mu}\eta_{\nu}=\partial_{\nu}\eta_{\mu},
\end{eqnarray}
and
\begin{eqnarray}
g=g_{\mu\nu}d\theta^{\mu}d\theta^{\nu}=g^{\alpha\beta}d\eta_{\alpha}d\eta_{\beta}=d\eta_{\alpha}d\theta^{\alpha},
\end{eqnarray}
where $g^{\alpha\beta}=\partial^{2}\psi/\partial\eta_{\alpha}\partial\eta_{\beta}$. For the Hessian potential given by Eq.~(\ref{fullpotential}), the Legendre parameters take the following forms:
\begin{eqnarray}
\eta_{1}=-\frac{\kappa}{f} \; , \; \eta_{i}=\frac{\kappa\theta^{i}}{f} \; (i=2,...,D).
\end{eqnarray}
The new parameter set $\by$ is defined by
\begin{eqnarray}
y^{1}=\sqrt{f} \; , \; 
y^{i}=\frac{1}{2}\theta^{i} \; (i=2,...,D). \label{yy}
\end{eqnarray}
Note that in this coordinates the Hessian potential is represented as
\begin{eqnarray}
\psi(\by)=-2\kappa\ln y^{1}.
\end{eqnarray}This means that the radial axis characterizes the magnitude of the entanglmenent entropy, and thus the entropy is a key factor of characterizing holographic renormalization.

Then, $\bt$ and $\mbox{\boldmath $\eta$}$ are represented by $\by$ as
\begin{eqnarray}
\theta^{1}=(y^{1})^{2}+2\sum_{i=2}^{D}(y^{i})^{2} \; , \; \theta^{i}=2y^{i} \; (i=2,...,D),
\end{eqnarray}
and
\begin{eqnarray}
\eta_{1}=-\frac{\kappa}{(y^{1})^{2}} \; , \; 
\eta_{i}=2\kappa\frac{y^{i}}{(y^{1})^{2}} \; (i=2,...,D).
\end{eqnarray}
The metric tensor is finally transformed into the Poincare disk representation of the hyperbolic geometry,
\begin{eqnarray}
g = d\eta_{1}d\theta^{1} + \sum_{i=2}^{D}d\eta_{i}d\theta^{i} = \frac{4\kappa}{(y^{1})^{2}}\left\{ (dy^{1})^{2} + \sum_{i=2}^{D}(dy^{i})^{2} \right\}.
\end{eqnarray}
Therefore, the Hessian potential of free fermoins is mapped onto the hyperbolic metric exactly. In this sense, the Fisher geometry can capture basic properties of the AdS/CFT correspondences. Note that the proper measure with physical units would be
\begin{eqnarray}
ds^{2}=4\kappa g,
\end{eqnarray}
since $\kappa$ is proportional to the curvature radius $l$ of AdS or the central charge $c$ of corresponding CFT~\cite{Brown}.

When we approach the boundary of Poicare disk $y^{1}\rightarrow 0$, the original subsystem size diverges $L\rightarrow\infty$. This clearly shows a kind of bulk/boundary-type correspondence in a sense that the UV limit of the original quantum state is located at the boundary of hyperbolic space.

\subsection{Consistency with Ryu-Takayanagi formula}

Tha Ryu-Takayanagi formula is a key method to calculate the entanglement entropy in the holographic side. In the previous works on entanglement thermodynamics, this formula is used as a kind of an indispensable dictionary to convert the entropy data into a geometric quantity by combining with the entropy-energy relation. This feature can be simply understandable by means of Hessian geometry. Let us look at this feature briefly. The Ryu-Takayanagi formula is represented as
\begin{eqnarray}
S=\frac{\gamma_{A}}{4G},
\end{eqnarray}
where $\gamma_{A}$ denotes the minimal surface area that surrounds the subsystem $A$, and $G$ is the Newton constant. In AdS${}_{2+1}$, a differential geometrical calculation tells us
\begin{eqnarray}
\gamma_{A} = 2l\ln L = -l\ln\theta^{1}, \label{RT}
\end{eqnarray}
with the curvature radius $l$. By combining Eq.~(\ref{RT}) with Eq.~(\ref{dS}), we find
\begin{eqnarray}
-\frac{l}{4G}\frac{1}{\theta^{1}}\simeq -\theta^{1}g_{11},
\end{eqnarray}
and then
\begin{eqnarray}
g_{11}\simeq\frac{l}{4G}\frac{1}{(\theta^{1})^{2}}=\partial_{1}\partial_{1}\left(-\frac{c}{6}\ln\theta^{1}\right)=\partial_{1}\partial_{1}\psi.
\end{eqnarray}
This is really consistent with necessary conditions $\kappa=c/6$ and $g_{11}=\partial_{1}\partial_{1}\psi$. The result means that the Ryu-Takayanagi formula combined with Brown-Henneaux central charge is equivalent to the present representation of the first law of entanglement thermodynamics.

\section{Relation between Hessian potential and entangling surface}

\subsection{Entropy as conserved information flow across the surface}

To understand geometric meaning of the Hessian potential seems to be a key for full construction of the theory of entanglement thermodynamics. In particular, we would like to focus on how the domain of the Hessian potential is related to an entangling surface. For this purpose, we go back to Eq.~(\ref{dS}) again. Multiplying $d\theta^{\nu}$ on both sides of Eq.~(\ref{dS}) and taking convention by the index $\nu$, we obtain
\begin{eqnarray}
\partial_{\nu}S(\bt)d\theta^{\nu}=-\theta^{\alpha}g_{\alpha\nu}(\bt)d\theta^{\nu}.
\end{eqnarray}
The left hand side of this equation is equal to $dS$. Thus, we can derive the derivative form of the first law,
\begin{eqnarray}
dS = -\theta^{\alpha}g_{\alpha\nu}(\bt)d\theta^{\nu} = -\theta^{\alpha}\partial_{\nu}\eta_{\alpha}d\theta^{\nu} = -\theta^{\alpha}d\eta_{\alpha}. \label{differentialform}
\end{eqnarray}
This is a sort of entropy-energy relation. Of course this covariant form originates in the exponential family form, and we guess that this is related to an invariant quantity associated with information flow in the classical side.

To examine this feature, we integrate Eq.~(\ref{differentialform}) over a closed hypersurface $\Sigma$, and the corresponding entanglement entropy $S_{\Sigma}$ may be described as
\begin{eqnarray}
S_{\Sigma}=-\int_{\Sigma}\theta^{\alpha}d\eta_{\alpha}=-\int_{\Psi}\nabla_{\alpha}\theta^{\alpha}d\Psi, \label{sigma}
\end{eqnarray}
where $\Sigma$ is the surface to which $d\eta_{\alpha}$ is a normal vector. Here, we have used the Gauss's law to derive the last equation in which the volume $\Psi$ is surrounded by the boundary $\Sigma$, i.e. $\Sigma=\partial\Psi$. Since the AdS space has the open boundary at $y^{1}\rightarrow 0$ in the Poincare disk representation, we should be careful about a fact that $\Sigma$ is in general different from the minimal surface $\gamma_{A}$ in the Ryu-Takayanagi formula. Thus, we may obtain an alternative and more global view for the area formula of the entanglement entropy in terms of holography. As is well known, this integral measures how much current flow occurs throughout the boundary $\Sigma$. In the present case, the canonical parameters can be regarded as the information flow, and their conjugate fields are the origin of the current flow.

According to the definition of the Gauss's law, we find
\begin{eqnarray}
d\eta_{\alpha} &=& n_{\alpha}d\Sigma , \\
\eta_{\alpha} &=& \partial_{\alpha}\psi , \\
\Sigma &=& \partial\Psi ,
\end{eqnarray}
where $n_{\alpha}$ is a unit vector normal to the surface $\Sigma$. Thus, $\Psi$ can almost be regarded as the Hessian potential $\psi$. Our conjecture is that the domain of the Hessian potential is related to the bulk area $\Psi$. Here, we would like to ask two questions: one is to prove the conjecture, and the other one is about how we can relate $\Sigma$ with a kind of entangling surface. However, the above statement is too naive, since $\psi$ is not a closed hypersurface. The potential function $\psi$ corresponds to a map from a $D$-dimensional vector to a scaler value, $\psi:R^{D}\rightarrow R$. The image is not doubled. Thus, we think that the domain boundary of $\psi$ seems to be a half of $\Psi$ owing to the positivity condition $\theta^{1}>0$. By pasting analytically connected region together, we may construct the full volume $\Psi$. We will mention how this kind of extention is naturally contained in our theory.

Before going into detail of the analysis of $\Sigma$, it is important to examine the meaning of $\nabla_{\alpha}\theta^{\alpha}$ in Eq.~(\ref{sigma}). Let us remember a conservation law for local vector current: $\nabla_{\alpha}q^{\alpha}=\partial_{\alpha}q^{\alpha}+\Gamma^{\alpha}_{\;\beta\alpha}q^{\beta}=0$. In the Hessian geometry, the Christoffel symbol is simply given by
\begin{eqnarray}
\Gamma^{\lambda}_{\;\mu\nu} = \frac{1}{2}g^{\lambda\tau}\left(\partial_{\mu}g_{\tau\nu}+\partial_{\nu}g_{\mu\tau}-\partial_{\tau}g_{\mu\nu}\right) = \frac{1}{2}g^{\lambda\tau}\partial_{\tau}\partial_{\mu}\partial_{\nu}\psi(\bt).
\end{eqnarray}
Then, we can evaluate $\nabla_{\alpha}\theta^{\alpha}$ as follows:
\begin{eqnarray}
\nabla_{\alpha}\theta^{\alpha} &=& D+\frac{1}{2}g^{\alpha\tau}\left(\partial_{\tau}\partial_{\alpha}\partial_{\beta}\psi(\bt)\right)\theta^{\beta} \nonumber \\
&=& D + \frac{1}{2}g^{\alpha\tau}\left\{\partial_{\tau}\left(\partial_{\alpha}\partial_{\beta}\psi(\bt)\theta^{\beta}\right)- g_{\alpha\tau}\right\} \nonumber \\
&=& \frac{1}{2}D-\frac{1}{2}g^{\alpha\tau}\partial_{\tau}\partial_{\alpha}S(\bt) \nonumber \\
&=& -\frac{1}{2}g^{\alpha\tau}\partial_{\tau}\partial_{\alpha}\left(S(\bt)-\psi(\bt)\right).
\end{eqnarray}
This result actually suggests that $\theta^{\alpha}$ is local conservation current and only the boundary term may survive after the volume integration over $\Psi$, since we can identify $\psi$ with $S$ after the second derivative. Now, we have assumed that we can defferentiate the potential as many as possible. However, this feature may violate at the domain boundary of the potential. This is an important point here, and is also an evidence of importance of the boundary $\Sigma$. The identification between $S$ and $\psi$ is a quite nontrivial thing special for the present entanglement case, and in this case the extensive feature of the entropy violates. We may relax the condition for the violation of extensive nature of the entropy, when $S$ and $\psi$ are not identified with each other but the decay of $\nabla_{\alpha}\theta^{\alpha}$ is quite rapid. These situations are completely different from standard thermodynamics. The present approach is basically parallel to the method of standard thermodynamics, but these exceptional features dominate the originality of the entanglement thermodynamics.

\subsection{Domain boundary of Hessian potential: emergence of alternative area formula}

To determine the shape of the hypersurface $\Sigma$, we again consider the 1D free fermion case. We transform the $\by$ representation of the Fisher metric into a new coordinate system slightly different from the original model parameters $(L,h,t)$. At first, we notice
\begin{eqnarray}
g=\frac{4\kappa}{(y^{1})^{2}}\left\{ (dy^{1})^{2}+(dy^{2})^{2}+(dy^{3})^{2} \right\}
\end{eqnarray}
and
\begin{eqnarray}
y^{1}=\frac{1}{L}\sqrt{1-\frac{1}{2}h^{2}-\frac{1}{2}t^{2}} \; , \; y^{2}=\frac{1}{2}h \; , \; y^{3}=\frac{1}{2}t.
\end{eqnarray}
With respect to this mapping, we introduce new coordinates $(L,H,T)$ as
\begin{eqnarray}
y^{1}=\frac{1}{L}\sqrt{1-\frac{1}{2}H^{2}-\frac{1}{2}T^{2}} \; , \; 
y^{2}=\frac{H}{\sqrt{2}L} \; , \; y^{3}=\frac{T}{\sqrt{2}L} .
\end{eqnarray}
Unfortunately, this is slightly different from the original coordinates $(L,h,t)$, but this minor change of coefficients is necessary to derive a correct form of the metric. Taking the polar coordinates
\begin{eqnarray}
H=\sqrt{2}r\cos\phi \; , \; T=\sqrt{2}r\sin\phi ,
\end{eqnarray}
we can derive
\begin{eqnarray}
g=\frac{4\kappa}{1-r^{2}}\left\{ (d \ln L)^{2} + \frac{dr^{2}}{1-r^{2}} + r^{2}d\phi^{2} \right\}, \label{g1}
\end{eqnarray}
where the boundary of the potential is characterized by $r=1$, and for this value the metric diverges.

\begin{figure}[htbp]
\vspace{2mm}
\begin{center}
\includegraphics[width=10cm]{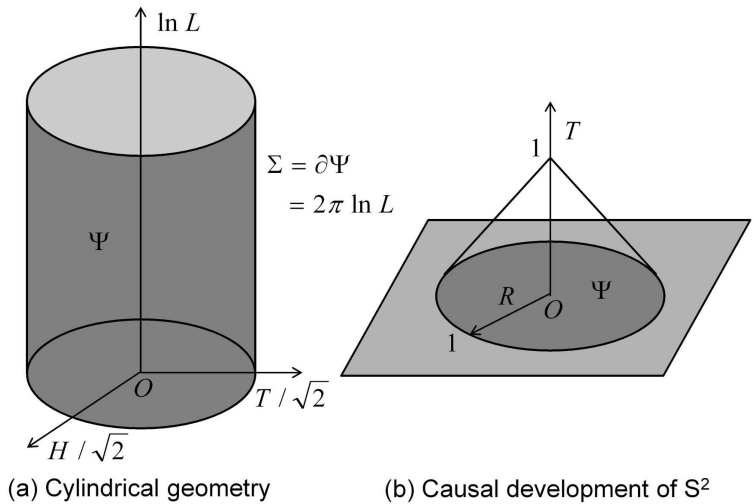}
\end{center}
\caption{(Color online) (a) A half of cylindrical geometry, and (b) Causal development of $S^{2}$. In Ref.~\cite{Casini}, these spaces are denoted as $\cal{H}$ and $\cal{D}$, respectively. However, it should be noted that $T$ is not real time in the present theory.}
\label{fig1}
\end{figure}

Equation~(\ref{g1}) corresponds to two copies of cylindrical geometry, $R_{+}\times S^{2}$ and $R_{-}\times S^{2}$, as shown in Fig.~\ref{fig1}(a). Since $L\ge 1$ for the discretized lattice size $L$, we find $\ln L\ge 0$ and this is the origin of $R_{+}$. In Eq.~(\ref{g1}), $d\ln L$ is squared. Thus if we define $\tau=\ln L$ and analytically connect the present theory to the negative $\tau$ region, we obtain full $R\times S^{2}$ geometry. This is the reason for the difference between $\Psi$ and $\psi$, as previously discussed.

To confirm the cylinder feature, we further take coordinate transformation as
\begin{eqnarray}
\tau=\ln L \; , \; r=\sin u ,
\end{eqnarray}
and we find
\begin{eqnarray}
g=(\Omega_{1})^{2}\left( d\tau^{2} + du^{2} + \sin^{2}u d\phi^{2} \right),
\end{eqnarray}
where the prefactor is given by
\begin{eqnarray}
(\Omega_{1})^{2}=\frac{4\kappa}{\cos^{2}u}.
\end{eqnarray}
We can eliminate the prefactor by an appropriate conformal transformation. To find the cylindrical feature more explicitely, it would be better to find coordinate transformation in which the cylindrical metric is represented by the flat Euclidean metric with some boundary condition. For this purpose, we suppose
\begin{eqnarray}
ds^{2}=dT^{2}+dR^{2}+R^{2}d\phi^{2}.
\end{eqnarray}
Here we introduce
\begin{eqnarray}
T=\frac{\sinh\tau}{\cosh\tau+\cos u} \; , \; R=\frac{\sin u}{\cosh\tau+\cos u},
\end{eqnarray}
and then we obtain
\begin{eqnarray}
ds^{2}=\frac{1}{(\cosh\tau+\cos u)^{2}}\left( d\tau^{2} + du^{2} + \sin^{2}u d\phi^{2} \right).
\end{eqnarray}
Thus
\begin{eqnarray}
g=(\Omega_{2})^{2}\left( dT^{2} + dR^{2} + R^{2}d\phi^{2} \right),
\end{eqnarray}
with
\begin{eqnarray}
(\Omega_{2})^{2}=(\Omega_{1})^{2}(\cosh\tau+\cos u)^{2}.
\end{eqnarray}
When we represent $T$ and $R$ by $L$, we have
\begin{eqnarray}
T=\frac{L-\frac{1}{L}}{L+\frac{1}{L}+2\cos u} \; , \; R=\frac{2\sin u}{L+\frac{1}{L}+2\cos u}.
\end{eqnarray}
Thus $T$ is bounded by $0\le T\le 1$ for $L\ge 1$. At $L=1$, the radial parameter $R$ takes $R\le 1$, and $R=0$ for $L\rightarrow\infty$ and $T=1$. Thus, the parameter region is cone-like structure. The parameter region is shown in Fig.~\ref{fig1}(b).

The above discussion enables us to notice that the surface $\Sigma$ corresponds to the domain boundary of the Hessian potential, and $d\eta_{\alpha}$ is normal to $\Sigma$. For Eq.~(\ref{g1}), the Hessian potential is given by
\begin{eqnarray}
\psi(\by)=-2\kappa\ln y^{1}=2\kappa\ln L - \kappa\ln\left(1-r^{2}\right).
\end{eqnarray}
The potential does not depend on $\phi$, and thus $d\eta_{\alpha}$ is perpendicular to the $\phi$ direction. Furthermore, the $L$ and $r$ dependences are decoupled. The normal direction is determined by the radial axis $r$ only. Thus, we realize that the boundary of the Hessian potential really corresponds to $\Sigma$, and $d\eta_{\alpha}$ is actually a normal vector to the hypersurface $\Sigma$.

It is interesting to notice that the boundary area is proportional to $\ln L$, and this seems to represent the magnitude of the entanglement entropy consistent with the scaling formula.
\begin{eqnarray}
\Sigma=\partial\Psi=2\pi\ln L,
\end{eqnarray}
where the prefactor $2\pi$ would be the entanglement temperature. This seems to be a new type of holographic entanglement entropy calculation alternative to the Ryu-Takayanagi formula. The main reason for the difference is that our coordinates are not usual spacetime but model parameters. However, it has been discussed in Ref.~\cite{Casini} which aims to derive the holographic entanglement entropy that the causal development of a CFT inside a spherical surface $S^{D-2}$ of Minkowski space $R^{1,D-1}$ can be mapped onto the thermal behavior in $R\times S^{D-1}$ (or $R\times H^{D-1}$). In this case also, the presence of local operators throughout coordinate transformation is a key ingredient. We believe that the emergence of this cylindrical geometry is not accidential, and capture the essential feature of the holographic entropy.

In 2D cases ($d=2$), $(d\ln L)^{2}$ in Eq.~(\ref{g1}) is replaced with $dL^{2}$ owing to Eq.~(\ref{2D}), and the geometry becomes $R\times S^{3}$. In this case, $\Sigma\propto L=L^{d-1}$, and the area-law scaling appears. Thus, our conjecture seems to be correct for higher-dimensional cases.

\subsection{Connection to Rindler wedge}

Finally, we briefly comment on a close relationship of the present approach to the physics of Rindler spacetime to confirm that $\Sigma$ actually corresponds to the entangling surface. We introduce the following transformation:
\begin{eqnarray}
\tanh\tau=\frac{\sin T}{\cosh U} \; , \; \tan u=\frac{\sinh U}{\cos T},
\end{eqnarray}
where $U\rightarrow\infty$ for $\tau=T=0$ and $u=\pi/2$. Then, we have
\begin{eqnarray}
d\tau^{2}+d u^{2} + \sin^{2}u d\phi^{2} = \frac{dT^{2}+dU^{2}+\sinh^{2}Ud\phi^{2}}{\cosh^{2}U-\sin^{2}T}.
\end{eqnarray}
This is further transformed into
\begin{eqnarray}
dU^{2}+\sinh^{2}Ud\phi^{2}=\frac{1}{z^{2}}\left(dz^{2}+dx^{2}\right),
\end{eqnarray}
with use of the following new coordinates
\begin{eqnarray}
x+iz=\rho e^{i\phi} \; , \; \cosh U=\frac{1+\rho^{2}}{1-\rho^{2}}.
\end{eqnarray}
Therefore, we find
\begin{eqnarray}
g = (\Omega_{3})^{2}\left\{ dT^{2}+\frac{1}{z^{2}}\left(dz^{2}+dx^{2}\right)\right\} = (\Omega_{4})^{2}\left(z^{2}dT^{2}+dz^{2}+dx^{2}\right), \label{Rindler}
\end{eqnarray}
where conformal factors are respectively defined by
\begin{eqnarray}
(\Omega_{3})^{2}=(\Omega_{1})^{2}\frac{1}{\cosh^{2}U-\sin^{2}T},
\end{eqnarray}
and
\begin{eqnarray}
(\Omega_{4})^{2}=(\Omega_{3})^{2}\frac{1}{z^{2}}.
\end{eqnarray}
They can all be removed by appropriate conformal transformations.

The metric of Eq.~(\ref{Rindler}) in the imaginary time $\tilde{T}=iT$ represents the Rindler wedge. Actually, when we take the null coordinates in the Rindler space as
\begin{eqnarray}
X^{\pm}=ze^{\pm \tilde{T}}, \label{Rindler}
\end{eqnarray}
and then we see
\begin{eqnarray}
g=(\Omega_{4})^{2}\left(dX^{+}dX^{-}+dx^{2}\right).
\end{eqnarray}
Here, we see $X^{+}+X^{-}=2z\cosh\tilde{T}>0$ and $X^{+}-X^{-}=2z\sinh\tilde{T}>0$. Remember that our $T$ coordinate is not time but is a kind of a length scale associated with the size $L$ of our partial system $A$. This feature coinsides with Eq.~(\ref{Rindler}). These results are also strong evidences that the domain structure of the Hessian potential determines the entangling surface in the holographic side.

\section{Summary}

We have examined the first law of entanglement thermodynamics by means of Hessian geometry. A key ingredient is the special functionality of the Hessian potential. The Hessian potential contains enough information of both the AdS/CFT correspondence and a kind of area law formula of the entanglement entropy as boundary flow of conserved current of information. This boundary flow of information is guaranteed by the identification between the Hessian potential and the entanglement entropy after the second derivative by the canonical parameters, and this feature is quite special for the entanglement case. This idenfication guarantees violation of extensive nature of the entanglement entropy. Our approach is basically consistent with Ryu-Takayanagi formula, but at the same time our area formula obtained here seems to be a new type. This is due to our special selection of coordinates that are not real spacetime. Further examination of the present achievement will shed new light on the physics of AdS/CFT correspondence and notions of holographic entanglement entropy.

\acknowledgments
I acknowledge Isao Maruyama, Kazuo Goroku, and Takumi Sasaki for useful discussions. This work was supported by JSPS KAKENHI Grant Number 15K05222.

\end{document}